# To save the Environmental degradation: Please pay for ecosystem services！

## Yijie Li, Jing Chen, Bernie Liu

**Abstract:** Project planners should consider the monetary value of ecosystem services in the regional service category and add it to the cost calculation of the project as environmental cost. Managers should monitor and regularly maintain the ecological environment within the project area to maintain the benefits of environmental costs.With regard to the problem of predicting how the model changes over time, we use LSTM to realize dynamic evaluation. The analysis that has been made shows that the monetary value of urban ecological services has increased. The environment is still being damaged. Through neural network memory and learning, the problem can be found, and the model can be modified in time.

# 1. Introduction

## 1.1 Background

In traditional land use projects, there is a lack of protection of the ecological environment, ignoring the impact of land use on its ecosystem services (of services obtained by human beings from natural ecosystems that directly or indirectly support human production and life) and changing in ecosystem services during utilization. In every decision of land use, due to the qualitative thinking, people lack the consideration of impact of the ecological circle, which leads to the degradation of the environment in the cumulative changes. Therefore, to reduce the economic costs associated with inappropriate land use, we have created an ecosystem service assessment model to consider ecosystems. Understanding the environmental and economic costs of land use projects in maintaining the biosphere.

## 1.2 Our work

We need to establish an ecosystem services model to evaluate the capacity of ecosystem services and to estimate the economic costs of maintaining the environment in land-use projects. Provide cost-benefit analysis for different scale land development projects. The model will be applied to land use projects of different scales, and suggestions for improvement will be put forward after testing.

To address these problems, we will take the following steps：

**Make assumptions and marks.** By presenting assumptions, ignoring minor effects reduces the focus of dealing with the problem.

**Make notations.** List the important symbols in the paper and make clear the meaning of them to make clear the model.

**Enumerate the influencing factors.** List the relevant important factors in the urban context: population density, electromagnetic wave impact (complex communication network), carbon emissions from motor vehicles, light intensity (neon), climate conditions in the GDP, area of the region, weight determination, the value grade of ecological service in City L was determined.

## 2. Assumptions and Justification

In order to make our model more suitable for real life conditions, we make the following assumptions.

We assume that we do not focus on such factors as science, technology, culture and policy, which will have a significant impact on the local community.

We assume that there will be no major natural disasters and other accidents in the project area, and that the international situation is relatively stable.

# 3. Notations

We show some of the symbols used in this article in Table 3-1

| Symbols | Definition |
|---------|------------|
| S | Entropy value of system |
| R | Evaluation matrix |
| W | Weight matrix |
| P | Monetary value of Urban Ecological Services |
| σ | Coefficient of viscosity (0 ＜σ＜1） |
| f | Forgotten door |
| i | In-gate |
| c | Cell state |
| o | Out-gate |
| b | Partial roof |

Table 3-1 the symbols used in the article

# 4. Development of Urban ecosystem Service Evaluation based on InVEST Model

## 4.1 Summarize

The Integrated Assessment Model of ecosystem Services (InVEST) is the most widely used and powerful assessment model in the field of ecosystem service assessment. By simulating different land use scenarios, the model sets the necessary data and obtains the value of ecosystem services, which is presented visually and intuitively. These are the unique advantages of the InVEST model.

The InVEST model covers a variety of the assessment modules of ecosystem services, currently including the ecosystem assessment of freshwater, marine and terrestrial, and has been applied in the protection of lakes, forests and the assessment of coastal ecosystem services.

However, there is not a single module in the InVEST model to evaluate the ecological services in urban areas and crowd gathering areas. With the acceleration of urbanization in some countries, the demand of the assessment of urban ecosystem service is growing. Therefore, this paper hopes to develop the evaluation module of urban ecosystem services based on InVEST model, which can provide the basis for the decision of urban land use projects.

## 4.2 The assessment module of urban ecological services

### 4.2.1 Index of evaluation system

First, we need to identify evaluation factors tha should play a key role in urban ecosystem. From the following five general elements: urban economic development, population correlation and distribution, ecosystem resilience, urban land use, infrastructure construction, we construct the evaluation system of urban ecological services.

● Target layer: gain the weight of factors.

● Factors: urban economic development, population correlation and distribution, ecosystem resilience, urban land use, infrastructure construction.

● According to each of the generalized factors, continue to expand to divide the factors into smaller sub-factors, forming a sub-factor layer.

(1) Urban economic development level: the proportion of per capita GDP, information industry to GDP, tourism income to GDP, annual GDP growth rate.

(2) Population correlation and distribution: population density, natural population growth rate, average education time of population, proportion of aging population.

(3) Ecosystem resilience: the comprehensive utilization ratio of industrial solid waste, the ratio of government environmental protection investment to GDP, the comprehensive air pollution index, the treatment rate of urban life pollution.

(4) Urban land use: per capita road area, per capita green area, urban per capita housing area, per capita working area.

(5) Infrastructure construction: hydro-power coverage, transportation development, network communication coverage, urban green coverage.

### 4.2.2 The criteria of ecosystem services value

According to the relevant official statistics and rational use of our urban context, the value of ecosystem services from high to low, the evaluation criteria are: excellent, op, middle, low, very low five levels.

| Essential Factor | Sub-element | Unit | Evaluation Grade Division | | | | |
|---|---|---|---|---|---|---|---|
| | | | Excellent | Top | Middle | Low | Very low |
| City Economic Development | Per capita GDP | Ten thousand dollars | 14~20 | 6~14 | 4~6 | 0.6~4 | <0.6 |
| | Proportion of GDP in Information Industry | % | 40~60 | 25~40 | 10~25 | 5~10 | <5 |
| | GDP share of tourism income | % | 15~20 | 10~18 | 5~10 | 1~5 | <1 |
| | Annual GDP growth rate | % | 10~15 | 8~10 | 6~8 | 2~6 | <2 |

Table 4-1 Grade Division Data of City Economic Development

| Essential Factor | Sub-element | Unit | Evaluation Grade Division | | | | |
|---|---|---|---|---|---|---|---|
| | | | Excellent | Top | Middle | Low | Very low |
| Population Correlation And Distribution | Density of population | Ten thousand people/km2 | <1 | 1~2 | 2~3 | 3~4 | >4 |
| | Natural population growth rate | % | <1.0 | 1.0~2.0 | 2.0~4.5 | 4.5~5.0 | >5.0 |
| | Average length of education of the population | a | >13 | 10~13 | 8~10 | 4~8 | <4 |
| | Proportion of ageing population | % | <4 | 4~6 | 6~8 | 8~12 | >12 |

Table 4-2 Grade Division Data of Population Correlation and Distribution

| Essential Factor | Sub-element | Unit | Evaluation Grade Division | | | | |
|---|---|---|---|---|---|---|---|
| | | | Excellent | Top | Middle | Low | Very low |
| Ecosystem Resilience | Comprehensive utilization ratio of industrial waste | % | >95 | 70~95 | 50~70 | 30~50 | <30 |
| | Natural population growth rate | % | >5 | 2.98~5 | 1.12~2.98 | 0.85~1.12 | <0.85 |
| | Average length of education of the population | | >6 | 4~6 | 2~4 | 0.5~2 | <0.5 |
| | Proportion of ageing population | % | >80 | 60~80 | 40~60 | 20~40 | <20 |

Table 4-3 Grade Division Data of Ecosystem Resilience

| Essential Factor | Sub-element | Unit | Evaluation Grade Division | | | | |
|---|---|---|---|---|---|---|---|
| | | | Excellent | Top | Middle | Low | Very low |
| Urban land use | Road area per capita | ㎡/person | >30 | 20~30 | 15~20 | 10~15 | <10 |
| | Per capita green area | ㎡/person | >18 | 11~18 | 5~11 | 3~5 | <3 |
| | City per capita housing area | ㎡/person | >30 | 20~30 | 15~20 | 10~15 | <10 |
| | Per capita working area | ㎡/person | >30 | 15~30 | 10~15 | 6~10 | <6 |

Table 4-4 Grade Division Data of Urban land use

| Essential Factor | Sub-element | Unit | Evaluation Grade Division | | | | |
|---|---|---|---|---|---|---|---|
| | | | Excellent | Top | Middle | Low | Very low |
| Infrastructure Construction | Hydropower supply coverage | % | 100 | 95~100 | 90~95 | 80~90 | <80 |
| | Traffic perfection degree | % | >95 | 90~95 | 80~90 | 70~80 | <70 |
| | Network communication coverage | % | >99 | 95~99 | 90~95 | 70~90 | <70 |
| | City green coverage | % | >60 | 50~60 | 40~50 | 30~40 | <30 |

Table 4-5 Grade Division Data of Infrastructure Construction

### 4.2.3 Calculation of the weight of each factor

We use the entropy weight method to calculate the weight.

● Entropy weight method is a weight determination algorithm for objective assignment. The algorithm process is as follows: information entropy -> entropy weight of each index -> revision -> objective weight value of each index weight.

● If the probability of each different state in the system is, then the entropy of the system is defined by the following formula:

$$s = -\sum_{i=1}^{N} P_i \times \ln(P_i) \qquad (4\text{-}1)$$

● Assume that there are M states, N indicators, primitive evaluation matrix:

$$R = \begin{pmatrix} r_{11} & r_{12} & ... & r_{1N} \\ r_{21} & r_{22} & ... & r_{2N} \\ ... & ... & ... & ... \\ r_{M1} & r_{M2} & ... & r_{MN} \end{pmatrix}_{M \times N}$$ (4-2)

It is the evaluation value of item I under the j th index.

● Solving the weight of each index

(1) Calculate the indicator weight of item i under indicator k:

$$p_{ik} = \frac{r_{ik}}{\sum_{i=1}^{M} r_{ik}}$$ (4-3)

(2) Calculate the entropy of the k th index:

$$s_k = -g \sum_{i=1}^{M} p_{ik} \times \ln(p_{ik}), g = \frac{1}{\ln M}$$ (4-4)

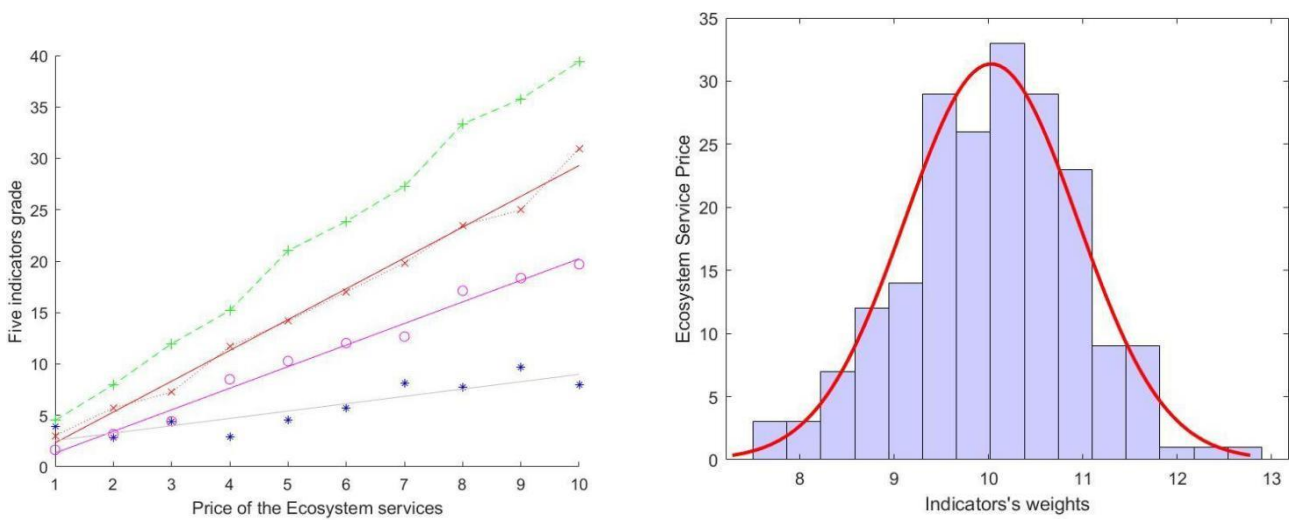

Fig 4-6 Fitting correlation curve of five elements & strip fitting

(3) Determine the combined weight of the indicator

$$w_k = \frac{(1-s_k)}{\sum\limits_{k=1}^{N}(1-s_k)} \qquad (4\text{-}5)$$

(4) The initial index weights given by an evaluation unit are expressed as By combining it with the entropy weight of the corresponding index, the comprehensive index weight value can be obtained:

$$\partial_k = \frac{\gamma_k w_k}{\sum\limits_{k=1}^{M}\gamma_k w_k} \qquad (4\text{-}6)$$

# 5. Assessment of Urban Ecological Service value used in Urban Rail Transit Project

## 5.1 Introduction to selected cities

The city we choose can be expressed as City L, which is a coastal city with a length of 67.85 km from east to west, a width of 37.5km from north to south, and a length of 103.3 km from the coast. The city's economic development, employment and civic welfare depends on the products and services provided by the marine and coastal ecosystems. For instance, abundant species and genetic resources, nutrient storage and recycling, purification of land-based pollutants, stabilization of shore lines, etc. At the same time, marine ecosystem plays a key role in regulating climate and maintaining air quality, which is the main carbon and oxygen source.

## 5.2 Urban Rail Transit Project

The city's new urban rail transit project, which revolves around the entire urban area, is in line with the substance of large-scale land use projects. As City L is a coastal city, it has both terrestrial and marine ecological environment. Therefore, combining the results of marine and coastal ecosystem value assessment with the urban ecosystem value assessment, the comprehensive value of ecosystem services of the urban rail transit project is obtained. And the cost-benefit ratio analysis before and after the environmental cost is added to obtain the true comprehensive evaluation of the project.

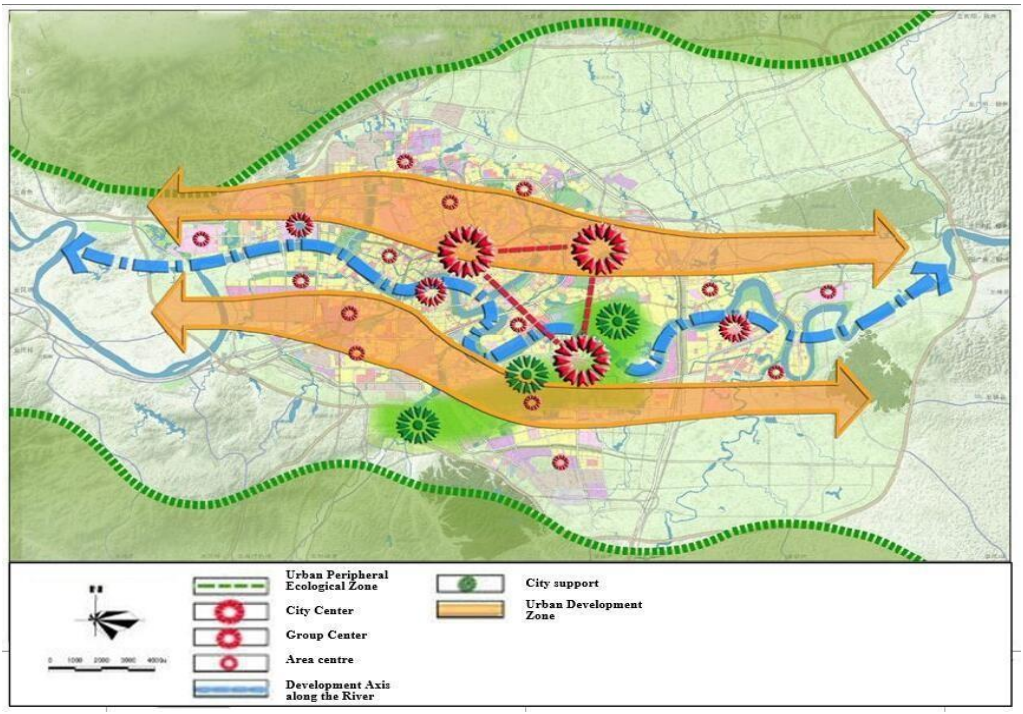

Fig 5-1 Urban rail transit hierarchical map

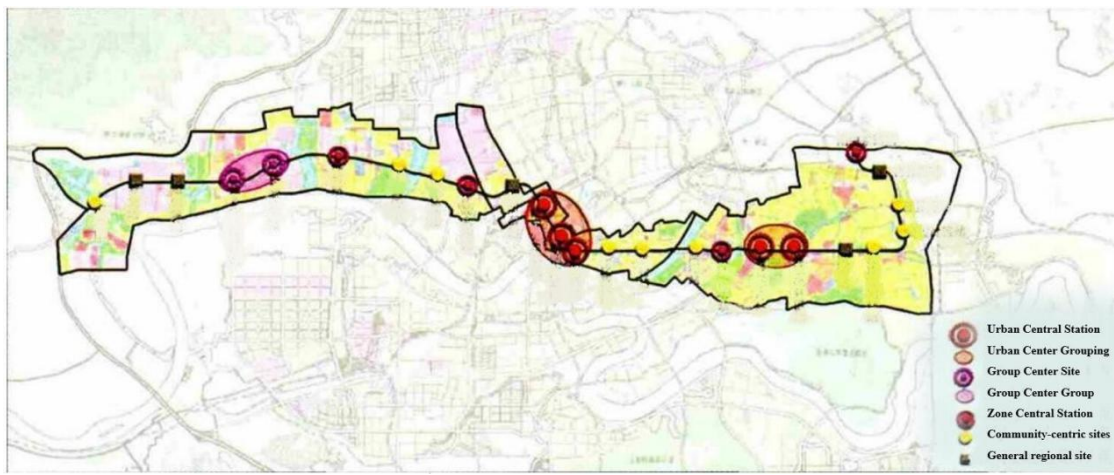

Fig 5-2 Urban rail transit process map

### 5.2.1 Assessment of marine coastal ecosystems

● Climate regulation service: the economic value of climate regulation service provided by the unit area study area of Pa. The cost of fixed $CO_2$ is recorded as $Cost_1$, and the cost of releasing $CO_2$ is $Cost_2$. The evaluation formula is as follows:

$$P_a = (1.63Cost_1 + 1.19Cost_2) \tag{5-1}$$

Calculated by InVEST software, the service value of climate control in City L is 0. 02($/m^2 \bullet a$).

● Pollution treatment and control services: the value of pollution treatment and control services per unit area is: Pev. The annual environmental capacity of the first species of pollutants in a sea area is $Xi(ton/a)$; the treatment cost of i pollutants is

Ci($/ton); The sea area is S(m²); The value of h (m) 's annual pollution treatment and control services is Pv($/m²•a). The evaluation formula is as follows:

$$\Delta v = \sum_{i=1}^{n} X_i C_i / Q$$

$$P_v = \Delta_v (Sh)$$

$$Pev = \frac{P_v}{S} = h \sum_{i=1}^{i=n} x_i c_i / Q$$

(5-2)

Calculated by InVEST software, the service value of pollution control ecosystem is 0.60($/m²•a).

● Landscape service: set up an index Uij representing the importance of the i region; Denote the j activity or the use of the j landscape in the i region (1 for use and 0 for non-utilization). The evaluation formula is as follows:

$$IS_i = \sum_{ij} U_{ij} I_j$$

(5-3)

The value of landscape ecosystem services calculated by InVEST software is 0. 11($/m²•a).

● Service of fishery resources: the value of serving fishery resources Pmf($/m²•a); The annual cost of marine fishing Cmf($/m²•a); The area of marine fishing S(m²). The evaluation formula is as follows:

$$P_{mf} = \frac{R_{mf} - C_{mf}}{S}$$

(5-4)

Calculated by InVEST software, the average value of ecosystem services for fishery resources is 0.32($/m²•a).

### 5.2.2 Assessment of urban ecosystems

Through the weights determined by the entropy weight method introduced in the previous section, we derive the calculation formula for the assessment of urban ecosystem services:

$$P_{urban} = \frac{P_{v\bullet}}{S} \frac{P_0 E}{\delta} \sigma P_s \rho$$

(5-5)

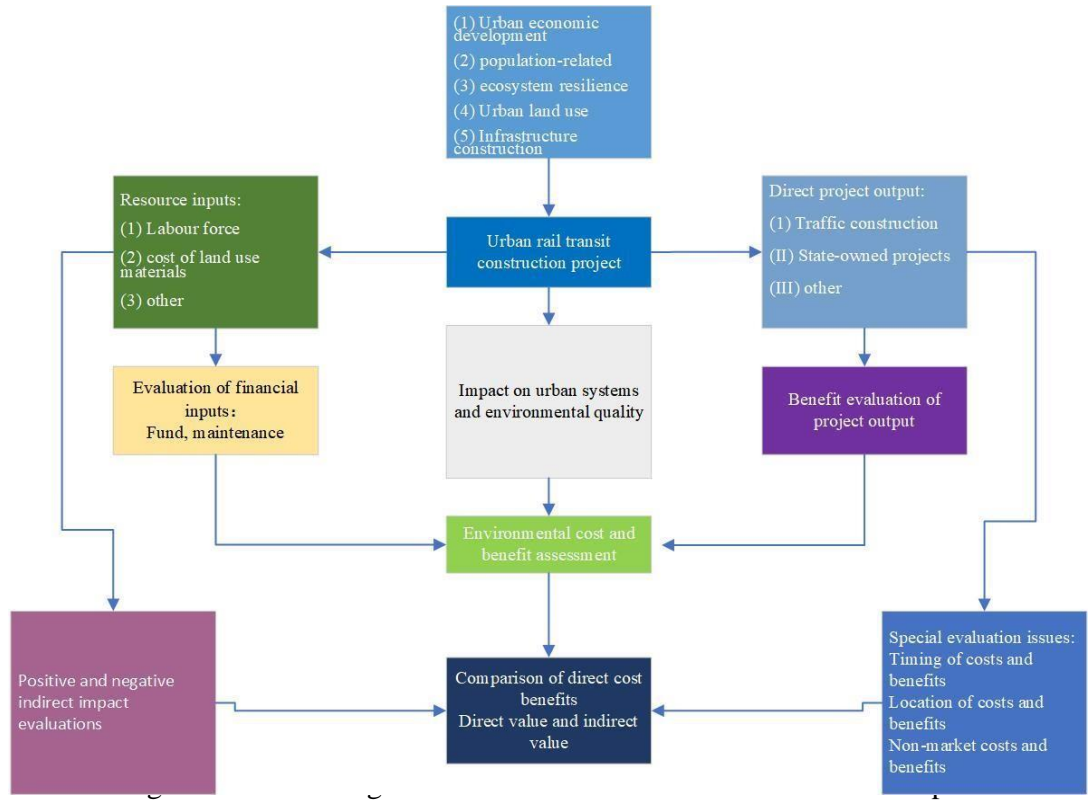

Fig 5-3 Urban ecological service assessment model establishment process

● $\rho$ is the level of ecological value assessed by entropy weight method and fuzzy evaluation model, which is equal to the specific monetary value of other ecological value evaluation in InVEST model.

(1) Using Fuzzy Comprehensive Evaluation to determine Evaluation Formula:

$$\theta = W \bullet R_{M \times N} \qquad (5\text{-}6)$$

$\theta$ is the result of the model evaluation, and W = (w₁,w₂,w₃,w₄,w₅) is the five general urban evaluation indicators (urban economic development, population correlation and distribution, ecosystem resilience, etc.) mentioned in the previous section. The comprehensive weight of urban land use and infrastructure construction, R is the relation matrix of subjection degree of each evaluation index to the standard matching.

$$R = \begin{pmatrix} R_{11} & R_{12} & R_{13} & R_{14} & R_{15} \\ R_{21} & R_{22} & R_{23} & R_{24} & R_{25} \\ R_{31} & R_{32} & R_{33} & R_{34} & R_{35} \\ R_{41} & R_{42} & R_{43} & R_{44} & R_{45} \\ R_{51} & R_{52} & R_{53} & R_{54} & R_{55} \end{pmatrix} \qquad (5\text{-}7)$$

(2) Concrete example calculation

By introducing the latest data from City L in 2019 into the evaluation model, the W and R matrix is obtained as follows:

$$R = \begin{pmatrix} 0.7723 & 0.5383 & 0.5443 & 0.032 & 0.733 \\ 0.0024 & 0 & 0.7742 & 0.042 & 0.134 \\ 0.8932 & 0.2234 & 0.2574 & 0.045 & 0.356 \\ 0.3334 & 0.1595 & 0.1241 & 0.024 & 0.251 \\ 0.1672 & 0.4325 & 0.0004 & 0.234 & 0.001 \end{pmatrix} \quad (5\text{-}8)$$

$$W = \begin{pmatrix} 0.452 & 0.675 & 0.986 & 0.463 & 0.523 \end{pmatrix} \quad (5\text{-}9)$$

$$\theta = W \bullet R = 0.5637 \quad (5\text{-}10)$$

$$P_{urban} = \frac{P_{v}}{S} \bullet \frac{P_0 E}{S} \bullet \sigma P^s \rho$$

$$= \frac{3.897}{\underline{\quad}} \bullet \frac{0.235 \bullet 10^s}{\underline{\quad\quad}} \bullet 0.0721 \bullet 2.232 \bullet 0.542 \quad (5\text{-}11)$$

$$0.324 \qquad 3.213$$
$$= 0.56(\$/m^2 \bullet a)$$

E is the original related cost of environmental protection in urban planning, $\sigma$ is viscosity coefficient ($0 < \sigma < 1$). $P_0$ is the comprehensive productivity of urban land per unit area.

**5.2.3  Analysis of Project Cost-benefit by adding Environmental cost**

**5.2.3.1  Cost analysis**

● **Determining cost unit**

Our team divided the cost into two categories: tangible cost and intangible cost.

● **Tangible cost ——** Tangible costs involved in the construction of urban rail transit

Ⅰ. Expenditures on investment-related commodities and raw materials.

Ⅱ. Operating expenses of urban rail transit.

Ⅲ. Real estate cost.

Ⅳ. Insurance, wages and taxes paid of employee.

Ⅴ. Maintenance of electricity and water charges.

   **Intangible cost**

Ⅰ. Time spent on projects (benefits from other projects spent on dunes).

Ⅱ. External energy for project operation.

Ⅲ. Assessment of possible business losses during project operation.

Ⅳ. Project environmental costs resulting from the value assessment of ecosystem services.

**5.2.3.2 Benefit analysis**

**Direct value:** it facilitates people's transportation and improves people's demand efficiency; urban rail transit construction can also increase tax revenue and contribute to the country.

**Indirect value:** taking the function provided by the ecosystem in the City as an example

**The value of choice:** the construction along the urban rail transit can be regarded as a new type of value, such as the landscape and tourist areas established along the line will benefit more because of the urban rail transit.

**5.2.3.3 Conclusion**

The cost benefit ratio of urban rail transit project before and after the addition of environmental cost was calculated by analyzing the cost and benefit of the urban rail transit project.

It is calculated that the cost-benefit ratio of the project before adding environmental cost is 0.583, and that after adding environmental cost is 0.696. Although the original cost increases after adding environmental cost, the overall benefit is increased. This reflects the need to add environmental costs to the total cost of the project.